\documentclass[10pt,letterpaper]{article}
\usepackage{opex3,array}
\usepackage{color}

\usepackage{amsmath,amssymb,graphicx}
\usepackage{multirow}

\usepackage{hyperref}
\usepackage{breakurl}
\usepackage{graphicx}

\newcommand{\e}{\eqref}

\begin{document}

\title{Nonlinear combining of multiple laser beams in Kerr medium}

\author{Pavel M. Lushnikov and Natalia Vladimirova}

\address{Department of Mathematics and Statistics, University of
New Mexico, USA}

\email{plushnik@math.unm.edu}

%\homepage{http://math.unm.edu/\textasciitilde% $\sim$ %\~{ } %\textasciitilde
% plushnik/} %% author's URL, if desired

\begin{abstract}
We consider combining of multiple laser beams into a single
near diffraction-limited beam by beam self-focusing (collapse) in a
Kerr medium. The beams with the total power above critical are
combined in the near field and then propagated in the Kerr medium.
Nonlinearity results in self-focusing event, combining multiple beams
into nearly a diffraction-limited beam that carries the critical power.
Beam quality of the combined beam is analyzed as a function of
the number of combining beams and the level of random fluctuations of the
combining beams phases.
\end{abstract}


%%%%%%%%%%%%%%%%%%%%%%% References %%%%%%%%%%%%%%%%%%%%%%%%%
%\bibliographystyle{osajnl}

%\bibliography{lushnikov,biblionls,biblio}

\bibliographystyle{osajnl}

%
%\begin{thebibliography}{99}
%
%\bibitem{bib1}P. J. Harshman, T. K. Gustafson, and P. Kelley, ``Title of paper," J. Chem. Phys. {\bf 3}, (to be published).
%
%\bibitem{gallo99} K. Gallo and G. Assanto, ``All-optical diode based on second-harmonic generation in an asymmetric waveguide,'' \josab {\bf 16}(2), 267--269 (1999).
%
%\bibitem{Masters98a} B. R. Masters, ``Three-dimensional microscopic tomographic imagings of the cataract in a human lens in vivo,'' \opex {\bf 3}(9), 332--338 (1998).
%
%\bibitem{Oron03} D. Yelin,  D. Oron,  S. Thiberge,  E. Moses, and Y. Silberberg, ``Multiphoton plasmon-resonance microscopy,'' \opex {\bf 11}(12), 1385--1391 (2003).
%
%\bibitem{samplefig}
%B.~N.~Behnken, G.~Karunasiri, D.~R.~Chamberlin, P.~R.~Robrish, and
%J.~Faist, ``Real-time imaging using a 2.8~THz quantum cascade laser
%and uncooled infrared microbolometer camera,'' \ol \textbf{33}(5),
%440--442 (2008).
%
%\end{thebibliography}
%

\section{Introduction}

A quick increase of the output power of fiber lasers in the last two
and half
decades~\cite{RichardsonNilssonClarksonJOSAB2011,JaureguiLimpertAndreasTunnermannNaturePhotonics2013}
resulted in reaching $\sim 10$kW in 2009 for the diffraction-limited
beam~\cite{GapontsevFominAbramovConf2010}. Currently, 10-kW single
mode and up to 100-kW multimode continuous-wave commercial fiber
lasers are available~\cite{IPGphotonics}, although the beam quality is
not yet specified. The growth of output power since 2009 has been
mostly stagnated because of the encountered mode
instabilities~\cite{JaureguiLimpertAndreasTunnermannNaturePhotonics2013,JaureguiEtAlSPIE2015}.
Further increase of the total power of the diffraction-limited beam is
possible through the coherent beam
combining~\cite{FanEEEJSelTopicsQuantElec2005,RichardsonNilssonClarksonJOSAB2011},
where the phase of each laser beam is controlled, aiming to produce a
combined beam with a coherent phase. Beam combining has been
successfully demonstrated only for several beams. E.g.,
Ref.~\cite{RedmondEtAlOptLett2012} has achieved combining of five
$500$W laser beams into a 1.9kW Gaussian beam with a good beam
quality, $M^2=1.1$.

Nonlinearity is expected to be the key issue for further output power
scaling of coherent beam
combining~\cite{RichardsonNilssonClarksonJOSAB2011}.  In
Ref.~\cite{LushnikovVladimirovaOptLett2014} we proposed that instead
of trying to overcome this difficulty with nonlinearity, one can use
nonlinearity to our advantage. We proposed to combine multiple laser
beams with uncorrelated phases into a diffraction-limited beam using
strong self-focusing in a nonlinear waveguide. Difficulty of that
proposal is a significant variation of the distance to the collapse
and sensitivity to phase distribution of the laser beams at the
entrance to the waveguide. To mitigate this problem and to avoid
catastrophic collapse, which could damage the waveguide, several
approaches were proposed in
Ref.~\cite{LushnikovVladimirovaOptLett2014}.  Another option is to
work with multi-core
fiber~\cite{ShtyrinaRubenchikFedorukTuritsynPRA2015}, which reduces
the efficiency of nonlinear combining.

\begin{figure}
\begin{center}
\includegraphics[width=4.325in]{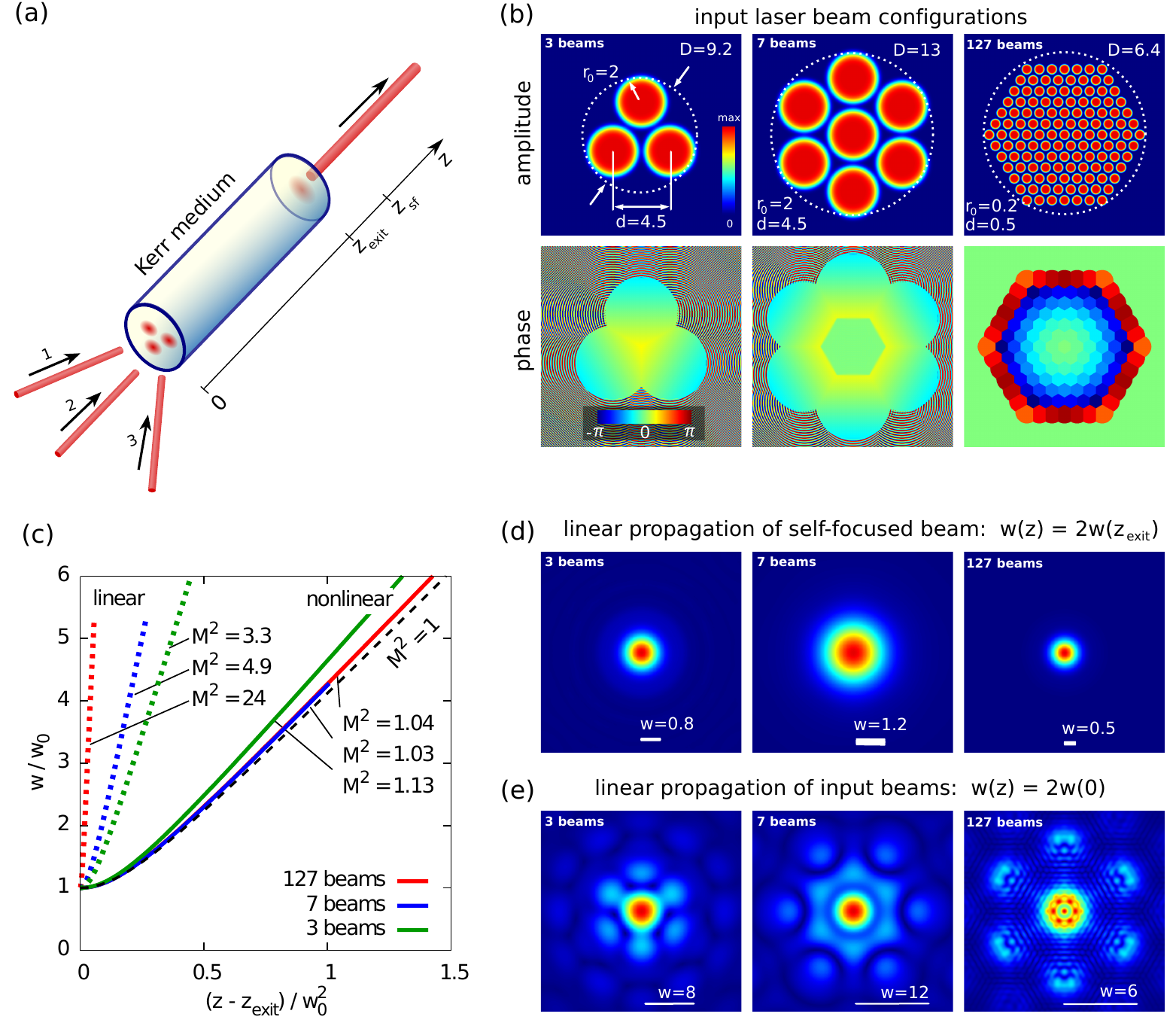}
\caption{
%
% NO LONGER THAN 5 LINES OF FORMATTED TEXT!
%
  (a) A schematic of  3-beam combining setup.
  %
  % The self-focusing interaction of beams in a slab of Kerr medium
  % results in a single beam exiting the medium.  This beam would
  %reach the nonlinear focus at $z_{\rm sf}$ if nonlinear medium were
  %extended.
  %
  % Here, $z_{sf}$ is the nonlinear focus distance provided beam
  % would propagate in Kerr medium until $z=z_{sf}.$
  (b) Amplitude and phase at the entrance surface for 3, 7 and 127 beams.
  %Positions of combining beams on the entrance surface of
  %entrance surface of nonlinear medium for 3, 7 and 127 beams.
  % here $d$ is the distance between the centers of neighboring beams.
  (c) $z$-dependence of the beam width $w$. Solid lines are for propagation of self-focused beams in vacuum after exiting the Kerr medium at $z=z_{\rm exit}$. Dotted lines of the same colors are for the propagation of   input beams  without Kerr medium ($z_{\rm exit}=0$ in that case). A black dashed line is for propagation of Gaussian beam in vacuum with the beam quality factor $M^2=1.$
  (d) A cross-section of amplitude of the self-focused beam propagating in vacuum for $z>z_{\rm exit}$ (corresponds to solid lines in (c)) at the
  location where $w(z) = 2w(z_{\rm exit})$.
  (e) A cross section of amplitude for beams propagating in vacuum   (corresponds to solid lines in (c) with $z_{\rm exit}=0$)  at the
  location where $w(z) = 2w(0).$ %
  }
\label{fig:multi_setup}
\end{center}
\end{figure}

In this paper we consider a different approach to nonlinear beams
combining that uses a slab of Kerr media instead of a waveguide.  We
now require that the phases of input beams  are close to each
other. The number of laser beams can vary significantly, as long as
the total power exceeds the critical power of self-focusing.  (In our
simulations we used $3$, $7$ and $127$ beams, but their number is
straightforward to increase.)  Figure~\ref{fig:multi_setup}a shows a
schematic of the nonlinear beam combining in a slab of Kerr medium.

The paper is organized as follows. In Section \ref{sec:Basicequations}
simulation settings are discussed together with the basic facts about
collapses.  Section \ref{sec:beamquality} considers the beam quality
$M^2$ of the combined beam and its sensitivity to the fluctuations of
the phases of combining beams.  In Section \ref{sec:Conclusion} the
main results of the paper are discussed.

\section{Basic equations and simulation parameters}
\label{sec:Basicequations}

We assume  that the pulse duration is long enough to neglect
time-dependent effects. (We estimate the range of allowed pulse
durations below.) The propagation of a quasi-monochromatic  beam
with a  single polarization through a Kerr media is given by
the nonlinear Schr\"odinger  equation (NLSE) (see e.g.
\cite{BergeSkupinNuterKasparianWolfBergePhysRep2007}):
\begin{eqnarray}\label{nlsdimensionall}
  i\partial_z\psi+\frac{1}{2k}\nabla^2\psi+\frac{kn_{2}}{n_0}|\psi|^{2}
  \psi=
  0,
\end{eqnarray}
where the beam is propagating along $z$-axis, ${\bf r}\equiv (x,y)$ are
the transverse coordinates, $\psi({\bf r},z)$ is the envelope of the
electric field,  $\nabla\equiv\left ( \frac{\partial }{\partial x},
\frac{\partial}{\partial y}\right )$, $k=2\pi n_0/\lambda_0$ is the
wavenumber in a medium, $\lambda_0$ is the wavelength in the vacuum, $n_0$ is
the linear index of refraction, and $n_2$ is the nonlinear Kerr
index.  The index of refraction is $n=n_0+n_2 I$, where $I=|
\psi|^2$ is the light intensity. In fused silica $n_0=1.4535$,
$n_2=3.2\cdot 10^{-16}\text{cm}^2/\text{W}$ for
$\lambda_0=790\text{nm}$ and
 $n_0=1.4496$,  $n_2=2.46\cdot
10^{-16}\text{cm}^2/\text{W}$ for $\lambda_0=1070\text{nm}$.
We bring NLSE \eqref{nlsdimensionall} to the dimensionless
form
\begin{eqnarray}\label{nls1}
  i\partial_z\psi+\nabla^2\psi+|\psi|^{2}\psi=
  0,
\end{eqnarray}
by the scaling transformation $(x,y)\to (x,y)w_0 $,   $z\to
2zkw_0^2$ and $\psi\to \psi n_0^{1/2}/(2k^2 w_0^2 n_2)^{1/2}$, where
$w_0$ is of the order of the waist of each input laser beam.

We also assume that the diameter of input beams in physical units is
large enough for the applicability of NLSE \eqref{nls1} (it is
sufficient for it to exceed several microns).  All different cases of
physical parameters considered below well satisfy this applicability
condition.

NLSE \eqref{nlsdimensionall} describes the catastrophic
self-focusing (collapse) of a the
laser beam
~\cite{VlasovPetrishchevTalanovRdiofiz1971,ZakharovJETP1972},
provided the power $P$ trapped in the collapse exceeds the critical value
\begin{align} \label{Pcdef}
P_c=\frac{N_c
  \lambda_0^2}{8\pi^2n_2n_0}\simeq\frac{11.70\,\lambda_0^2}{8\pi^2n_2n_0}.
\end{align}
Here,
%\begin{equation}\label{Ncdef}
$N_c\equiv 2\pi\int R^2 r dr=11.7008965\ldots$
%\end{equation}
is the critical power for NLSE \eqref{nls1} in dimensionless units and
$R(r)$, $r\equiv|{\bf r}|$ is the radially symmetric Townes
soliton~\cite{SulemSulem1999}. %\cite{ChiaoGarmireTownesPRL1964}
The Townes soliton is
defined as the ground state soliton of NLSE, $\psi=e^{i z}R(r)$, that is the solution of
\begin{equation} \label{Townes0}
-R+\nabla ^2 R+R^3=0
\end{equation}
centered at ${\bf r}=0.$ In fused silica,
$P_c\simeq \text{2MW}$  for $\lambda_0=790\text{nm}$ and  $P_c\simeq
\text{4.7MW}$ for $\lambda_0=1070\text{nm}$. In air, $P_c\simeq
\text{2.4GW}$  for $\lambda_0=790\text{nm}.$

Assume that $N$ linearly polarized laser beams enter a Kerr medium
at $z=0$, as shown in Fig. \ref{fig:multi_setup}.  (We note that
generalization to a case of arbitrary polarization is possible but is
beyond the scope of this paper.)  The initial condition for NLSE
\eqref{nls1} is the superposition of the super-Gaussian beams,
$\psi({\bf r},z)|_{z=0} = \sum_{n=1}^N \psi_n$, with $\psi_n = A_n
\exp\left(- \frac{({\bf r}-{\bf r}_n)^8}{r_0^8} + i\phi_n\right)$.
Here $r_0, \ A_n, \, \phi_n$ and ${\bf r}_n $ are the width, the
amplitude, the phase, and the location of the center of the $n$th
beam, respectively.  All beams have the same amplitude, $A=A_n.$ In
simulations with 3 or 7 beams, the beams of radius $r_0= 2$ are spaced
at the distance $d = 4.5$ between their centers, see
Fig.~\ref{fig:multi_setup}b. In 127-beam simulation, the beams with $r_0 =
0.2$ and spacing $d = 0.5$ are packed in the hexagon pattern. The
phases $\phi_n$ vary between simulations. For simulations with 3 and 7
beams, we use $\phi_n={\bf k}_{n,\perp} \cdot ({\bf r} - {\bf r}_n),$
where the transverse wavevectors ${\bf k}_{n,\perp}$ have the same
 absolute value, %amplitude
${\bf k}_{n,\perp} =k_\perp$, and are pointed to the origin
so that ${\bf k}_{n,\perp}=-k_\perp{\bf r}_n/ |{\bf r}_n|$. The only
exception is the beam at the center for $7$-beams case, where ${\bf
  k}_{n,\perp}={\bf 0}$. This phase distribution mimics the beams
of constant phase entering the media at converging angles. In
simulations with 127 beams, each beam has the plane wavefront with
the constant phase $\phi_n=- \frac{1}{2}\chi {\bf r}^2_n$ , so that
the collection of all beams imitates the quadratic phase of a single
pre-focused wide beam with the diameter $D$ schematically shown by
the dotted line in Fig.~\ref{fig:multi_setup}b.

To model nonlinear propagation, we use 4th order pseudospectral
split-step method with adaptive resolution up to $5120 \times 5120$
gridpoints in ${\bf r}$  and adaptive stepping in $z$.  Except for more
sophisticated initial conditions, the major difference with  earlier
simulations~\cite{LushnikovVladimirovaOptLett2014}  that we either
did nor used the waveguide or used waveguide (modeled by linear
potential in the circular form) of the diameter above $\gtrsim3$ of
the total initial diameter of all combined beams. Unless specified,
the size of each simulation was chosen to be large enough to avoid
influence of of waveguide or domain boundaries on the resulting
solutions.

Our choice of the combined power and distribution of all initial beams
ensures strong self-focusing along $z$ with the formation of the
high amplitude beam (the collapsing filament) centered at ${\bf r}=0$.
The collapse is well approximated by the rescaled Townes soliton
\cite{SulemSulem1999}:
\begin{eqnarray}\label{selfsimilar}
|\psi({\bf r},z)|\simeq{L(z)^{-1}}R(\rho), \quad \rho\equiv{r}/{L(z)},
\quad |{\bf r}|\equiv r,
\end{eqnarray}
where $L(z)$ is the $z$-dependent beam width. The explicit form of
$L(z)$ dependence was found in
Ref.~\cite{LushnikovDyachenkoVladimirovaNLSloglogPRA2013}, where it
was confirmed that collapses with Gaussian initial conditions agree
with $L(z)$ starting from the amplitude $|\psi|$ about 3-4 times above
the initial value.  In a similar way, the optical turbulence dominated
by collapses
\cite{LushnikovRosePRL2004,LushnikovRosePlasmPysContrFusion2006,LushnikovVladimirovaOptLett2010,ChungLushnikovPRE2011}
results in well defined filaments of the form \e{selfsimilar} as their
amplitudes exceed the background values in 3-4 times.  Combining of
multiple beams, however, delays the formation of the solution
\e{selfsimilar} and typically requires 5-10 fold increase of
initial amplitude to approach a diffraction-limited beam, as
detailed Table 1.

\section{Beam quality of the combined beam and effect of fluctuations  of initial beam phases}
\label{sec:beamquality}

We assume that after propagation through Kerr medium, the combined
beam exits the medium at $z=z_{\rm exit}$ and propagates in vacuum for
$z>z_{\rm exit}$.  Note that the beam exits nonlinear medium before it
would reach nonlinear focus at $z=z_{\rm sf}$ if nonlinear medium were extended, ie.e we always choose $z_{\rm
  exit}<z_{\rm sf}$, as shown in Fig.~1a.
We measure the beam quality factor $M^2$ by fitting the
effective width of the beam at $z>z_{\rm exit}$ with a quadratic
function~\cite{RossBookLaserBeamQualityMetrics2013},
\begin{equation}
  w^2(z) = w_0^2 + \left(\frac{2M^2}{k_0 w_0}\right)^2 (z-z_0)^2,
  \quad \text{where} \quad
 w(z)= \sqrt{ 2 \frac{\int |\psi|^2
      \,r^2 d^2{\bf r}}{\int |\psi|^2\, d^2{\bf r}} }.
\label{msq_fit}
%\label{w_def}
\end{equation}
Eq.~\e{msq_fit} with $M=1$ is the exact solution for the
propagation of a Gaussian beam in the vacuum. For non-Gaussian beams
$M^2 > 1$.  For general non-Gaussian beams the quantities $w_0$, $M^2$ and $z_0$ are determined from the fit as
follows.  At the first step, we choose a range of $z$ such that
$w^2(z)$ changes about twice compared with $w_0^2$ to determine $w_0$,
$M^2=M^2_{\rm short}$ and $z_0$.  At the second step, we use a range of
$z$ where $w^2(z)$ increases $\sim100$ times to fit for a new value of
$M^2=M^2_{\rm long}$, with $w_0$ and $z_0$ obtained in the first fit. This
 scheme is different from proposed
in~\cite{RossBookLaserBeamQualityMetrics2013} by using these two scales in $z$.
Note that if a significant noise is present in the system, then the
integration \e{msq_fit} can be done over areas where light intensity
$I=|\psi|^2$ exceeds a cutoff value, $|\psi|^2 \ge I_{\rm cutoff}$, as
suggested in Ref. \cite{RossBookLaserBeamQualityMetrics2013}.
However, we found that the cutoff introduces ambiguity in determining
$M^2$, so we choose to integrate in \e{msq_fit} over the entire
cross-section.  To ensure good quality of the beam,  we assumed that
a diaphragm that removes all components with $r>r_d=3L(z_{\rm exit}$)
is installed at $z=z_{\rm exit}$.  Here, $L(z)$ is determined by the
rescaled Townes soliton \e{Townes0}, \e{selfsimilar} as
$L(z)=R(0)/|\psi({\bf 0},z)|,$ where $R(0)=2.20620\ldots$
\cite{LushnikovDyachenkoVladimirovaNLSloglogPRA2013}. Note that the
diaphragm also removes the oscillating tails of Townes solition
\cite{LushnikovDyachenkoVladimirovaNLSloglogPRA2013}, so the power
of the filtered exit beam, $P_d$, is lower that the input power $P$ and can
fall below $P_c$.

Typical dependence $w(z)$ in vacuum, obtained using Eq.~\e{msq_fit},
is shown in Fig \ref{fig:multi_setup}c with $M^2\equiv M^2_{\rm long}$
and $P = 2P_c$.  The results of our simulations, summarized in Table~1,
demonstrate very good quality of the combined beams, while
propagation of the same input beams without self-focusing results in
very low beam quality as shown %by dotted lines
in Fig.~\ref{fig:multi_setup}c.
%
% Also $z_{\rm sf}$ in Fig.~\ref{fig:multi_setup}a schematically shows a nonlinear focus
% distance provided beam would propagate in Kerr medium until
% $z=z_{sf}$. Our choice was always $z_{\rm exit}<z_{\rm sf}.$
%
%
For a given input power, the increase of the pre-focusing parameter $\chi$ allows to
significantly reduce $z_{\rm exit}$ while keeping very good values of
$M^2$, although sometimes at the expense of reducing $P_d$.
%We have also
%obtained hight beam quality with input power as low as $1.6 P_c$. That
%power can be further reduced if Kerr waveguide is used instead of the
%slab of Kerr medium.
The input power can be reduced even further (as least to $1.6 P_c$)
if a Kerr waveguide is used instead of a slab of Kerr medium.

Last two columns of Table 1 provide physical values for the Kerr
medium thickness, $\tilde z_{\rm exit}$, and the input diameter,
$\tilde D$, of all combined beams, as shown in
Fig.~\ref{fig:multi_setup}b.  To find these values we use
$n_0=1.4496$, and $n_2=2.46\cdot 10^{-16}\text{cm}^2/\text{W}$ for
fused silica at $\lambda_0=1070\text{nm}$, which corresponds to the
wavelength of the commercially available 100kW continuous-wave (cw)
fiber laser \cite{IPGphotonics}.  We have also assumed that the
maximum of intensity at $z=z_{\rm exit}$ is $I=I_{\rm
  thresh}\simeq 5\cdot 10^{11}\text{W}/\text{cm}^2$.  The value for
$I_{\rm thresh}$ is obtained in experimental measurements of the
optical damage threshold in fused silica for $8\,\text{ns}$ pulses
\cite{SmithDoApplOpt2008}. If shorter pulses are considered then the
threshold is increased, e.g. $I_{\rm thresh}\sim 1.5\cdot
10^{12}\text{W}/\text{cm}^2$ for $14\text{ps}$ pulses
\cite{SmithDoApplOpt2008}, while cw operation requires $I_{\rm
thresh}\sim 10^{9}\text{W}/\text{cm}^2$ \cite{DawsonALLOptExpr2008}.
To use these alternative values, one need to re-evaluate the last
two columns of Table~1 according to scaling used in Eq.~\e{nls1}.
Note that neglecting contribution from a group velocity dispersion
in Eq.~\e{nls1} is justified for pulse durations above $\sim1$ps for
$z_{\rm exit}$ not exceeding several meters, as estimated in
Ref.~\cite{LushnikovVladimirovaOptLett2014}. Contributions of
stimulated Brillouin scattering and stimulated Raman scattering are
estimated in Ref.~\cite{LushnikovVladimirovaOptLett2014}.

%---------------------------------------------------------

\begin{table*}
\begin{tabular}{r | r | rrr | rrr | rrrrrrrrrrrr}
\hline
&  $k_\perp$ or $\chi$ & $\quad z_{\rm exit}$ &
$|\psi_{\rm exit}|$ & $|\psi_{\rm exit}/\psi_0|$ &
$M^2_{\rm short}$ & $M^2_{\rm long}$  &  $P_d/P_c$ &
$\tilde z_{\rm exit}$, cm & $\tilde D$, mm\\
\hline
%\multirow{6}{*}{3 beams}
 \parbox[t]{2mm}{\multirow{6}{*}{\rotatebox[origin=c]{90}{3 beams}}}
 & 0   &  4.10 &  4.90 &   5.43 &  1.189 & 1.183  & 1.0989  &  13.5 & 0.32\\
 & 0   &  4.25 &  9.98 &  11.07 &  1.119 & 1.119  & 1.0687  &  58.2 & 0.65\\
 & 0   &  4.29 & 20.43 &  22.66 &  1.114 & 1.114  & 1.0602  & 246.4 & 1.32\\
 & 0.3 &  2.00 &  5.31 &   5.89 &  1.201 & 1.194  & 1.1026  &   7.8 & 0.34\\
 & 0.3 &  2.10 &  9.12 &  10.11 &  1.134 & 1.134  & 1.0872  &  24.0 & 0.59\\
 & 0.3 &  2.14 & 17.07 &  18.94 &  1.131 & 1.131  & 1.0740  &  85.8 & 1.11\\
\hline
% 1.6 & 0   & 12.50 &  4.59 &   5.70 &  1.135 & 1.126  & 1.0077  &  36.2 & 0.30\\
% 1.6 & 0   & 12.80 &  9.82 &  12.18 &  1.086 & 1.086  & 1.0131  & 169.8 & 0.64\\
% 1.6 & 0   & 12.87 & 20.65 &  25.62 &  1.087 & 1.087  & 1.0125  & 755.2 & 1.34\\
% 1.6 & 0.3 &  3.00 &  4.78 &   5.93 &  1.102 & 1.098  & 0.9980  &   9.4 & 0.31\\
% 1.6 & 0.3 &  3.34 & 10.28 &  12.75 &  1.079 & 1.079  & 0.9952  &  48.6 & 0.67\\
% 1.6 & 0.3 &  3.45 & 21.47 &  26.63 &  1.078 & 1.078  & 0.9959  & 218.8 & 1.39\\
%\multirow{6}{*}{7 beams}
 \parbox[t]{2mm}{\multirow{6}{*}{\rotatebox[origin=c]{90}{7 beams}}}
 & 0   &  7.65 &   5.12 &   8.68 &  1.156  & 1.155 & 1.0448 &   28 & 0.60 \\
 & 0   &  7.82 &   9.85 &  16.69 &  1.109  & 1.109 & 1.0491 &  104 & 1.15 \\
 & 0   &  7.86 &  17.81 &  30.19 &  1.107  & 1.107 & 1.0408 &  343 & 2.08 \\
 & 0.3 &  5.15 &   5.02 &   8.51 &  1.034  & 1.034 & 1.0355 &   18 & 0.58 \\
 & 0.3 &  5.29 &  10.05 &  17.04 &  1.090  & 1.090 & 1.0176 &   74 & 1.18 \\
 & 0.3 &  5.34 &  18.77 &  31.82 &  1.090  & 1.090 & 1.0178 &  259 & 2.20 \\
\hline
%\multirow{6}{*}{127 beams}
\parbox[t]{2mm}{\multirow{9}{*}{\rotatebox[origin=c]{90}{127 beams}}}
& 0.5 & 0.373  &  6.95 &  5.01  & 2.065 & 1.858 & 0.9083 & 2.45 & 0.40 \\
& 0.5 & 0.435  & 11.17 &  8.05  & 1.038 & 1.037 & 0.9085 & 7.38 & 0.64 \\
& 1.0 & 0.196  &  6.94 &  5.00  & 1.941 & 1.796 & 0.9204 & 1.32 & 0.40 \\
& 1.0 & 0.220  & 11.15 &  8.04  & 1.592 & 1.572 & 0.8495 & 3.76 & 0.64 \\
& 1.0 & 0.231  & 13.75 &  9.91  & 1.141 & 1.135 & 0.8402 & 5.98 & 0.79 \\
& 2.0 & 0.106  &  6.93 &  5.00  & 1.526 & 1.516 & 0.6394 & 0.72 & 0.40 \\
& 2.0 & 0.115  & 11.02 &  7.94  & 1.417 & 1.412 & 0.6245 & 2.01 & 0.63 \\
& 2.0 & 0.122  & 13.85 &  9.98  & 1.363 & 1.359 & 0.6261 & 3.16 & 0.80 \\
\hline
\end{tabular}
\caption{
  Properties of self-focused beams formed by combining 3, 7, or 127
  beams with total input power $P = 2 P_c$: radial component of the
  wave vector, $k_\perp$ (for the 3-beam and 7-beam cases) or phase
  shift parameter $\chi$ (for the 127-beam case); maximum amplitude at
  the exit from nonlinear medium, $|\psi_{\rm exit}|\equiv|\psi({\bf
    0},z_{\rm exit})|$; ratio of exit amplitude to input amplitude,
  $|\psi_{\rm exit}/\psi_0|\equiv|\psi_{\rm exit}|/\max_{\bf
    r}|\psi({\bf r},0)|$; quality of the exit beam filtered by the
  diaphragm of radius $r_d = 3L(z_{\rm exit}$), measured at short
  distances $M^2_{\rm short}$ and at long distances $M^2_{\rm long}$;
  power of the exit beam filtered by the diaphragm, $P_d$; thickness
  of the slab of the medium, $\tilde z_{\rm exit}$, and total diameter of
  input beams, $\tilde D_{\rm }$, both in physical units.
  Here $\tilde z_{\rm exit}= (0.1376\, {\rm cm}) z_{\rm exit} |\psi_{\rm
    exit}|^2$ and $\tilde D= (0.0180\, {\rm mm})D |\psi_{\rm
    exit}|$ for the intensity maximum
  $I= 5\cdot 10^{11}\text{W}/\text{cm}^2$ at $z=z_{\rm exit}$.
  %
  % For 127-beam case, the second column is replaced by values of
  % $\chi$ which defines the phases of input beams as $\phi_n=-
  % \frac{1}{2}\chi {\bf r}^2_n$ , see Section \ref{sec:Basicequations}
  % for more details.
}
\label{table_all}
\end{table*}
%---------------------------------------------------------

We also studied a sensitivity of $M^2$ to $10\%$ random fluctuations
of input beams phases and found that $M^2$ changes by less that $1\%$,
 while the amplitude of the output beam is slightly reduced.
Thus, the parameters of Table~1 ensure that no catastrophic collapse is
possible for any random distribution of initial phases and
optical damage of Kerr medium can be avoided.

\section{Conclusion}
\label{sec:Conclusion}
In conclusion, we demonstrated the possibility of
nonlinear beam combining by propagating multiple laser beams in a Kerr
medium with realistic parameters. Table~1 shows that one can scale
the medium thickness $\tilde{z}_{\rm exit} $ down to $\sim1$cm  and the medium width down to
$\sim1$mm in fused silica with $\lambda_0=1070\text{nm}$. Further
improvements are possible if going beyond a flat distribution of
amplitudes of input beams. Strong self-focusing forms a single beam
of high quality with the power close to critical power of
self-focusing.  We  demonstrated that random fluctuations
$10\%$  of the beams phases do not alter the
obtained results. Note that if one allows the beams' phases to be
completely random, then instead of a slab of Kerr medium a
significantly longer waveguide is needed to achieve beam
combining~\cite{LushnikovVladimirovaOptLett2014}.

\section*{Acknowledgments}

This work was supported by NSF grant No. DMS-1412140.  Simulations
were performed at the Center for Advanced Research Computing, UNM, and
the Texas Advanced Computing Center using the Extreme Science and
Engineering Discovery Environment, which was supported by NSF Grant
No. ACI-1053575.  %


\begin{thebibliography}{10}
\newcommand{\enquote}[1]{``#1''}

\bibitem{RichardsonNilssonClarksonJOSAB2011}
D.~J. Richardson, J.~Nilsson, and W.~A. Clarkson, \enquote{High
power fiber
  lasers: current status and future perspectives,} J. Opt. Soc. Am. B
  \textbf{27}, B63--B92 (2010).

\bibitem{JaureguiLimpertAndreasTunnermannNaturePhotonics2013}
C.~Jauregui, J.~Limpert, and A.~T\"unnermann, \enquote{High-power
fibre
  lasers,} Nature Photonics \textbf{7}, 861 (2013).

\bibitem{GapontsevFominAbramovConf2010}
V.~Gapontsev, F.~A. Fomin, and M.~Abramov, \enquote{Diffraction
limited
  ultra-high-power fibre lasers,}  (2010). Paper AWA1 in Proc. Adv. Solid-State
  Photon. OSA Topical Meeting.

\bibitem{IPGphotonics}
\url{http://www.ipgphotonics.com} .

\bibitem{JaureguiEtAlSPIE2015}
C.~Jauregui, H.-J. Otto, N.~Modsching, J.~Limpert, and
A.~T\"unnermann,
  \enquote{Recent progress in the understanding of mode instabilities,} in
  \enquote{Proceedings of SPIE. Conference on Fiber Lasers XII - Technology,
  Systems, and Applications,} , vol. 9344, L.~Shaw and J.~Ballato, eds. (2015),
  vol. 9344, p. 93440J.

\bibitem{FanEEEJSelTopicsQuantElec2005}
T.~Y. Fan, \enquote{Experimental observations of the threshold-like
onset of
  mode instabilities in high power fibre amplifiers,} IEEE J. Sel. Topics in
  Quant. Elec. \textbf{11}, 567--577 (2005).

\bibitem{RedmondEtAlOptLett2012}
S.~M. Redmond, D.~J. Ripin, C.~X. Yu, S.~J. Augst, T.~Y. Fan, P.~A.
Thielen,
  J.~E. Rothenberg, and G.~D. Goodno, \enquote{Diffractive coherent combining
  of a 2.5 kw fiber laser array into a 1.9 kw gaussian beam,} Opt. Lett.
  \textbf{37}, 2832--2834 (2012).

\bibitem{LushnikovVladimirovaOptLett2014}
P.~M. Lushnikov and N.~Vladimirova, \enquote{Nonlinear combining of
laser
  beams,} Optics Letters \textbf{39}, 3429--3432 (2014).

\bibitem{ShtyrinaRubenchikFedorukTuritsynPRA2015}
O.~V. Shtyrina, A.~M. Rubenchik, M.~P. Fedoruk, and S.~K. Turitsyn,
  \enquote{Spatiotemporal optical bullets in two-dimensional fiber arrays and
  their stability,} Physical Review A \textbf{91}, 033810 (2015).

\bibitem{BergeSkupinNuterKasparianWolfBergePhysRep2007}
L.~Berg\'e, S.~Skupin, R.~Nuter, J.~Kasparian, and J.-P. Wolf,
  \enquote{Ultrashort filaments of light in weakly ionized, optically
  transparent media,} Rep. Prog. Phys. \textbf{70}, 1633–1713 (2007).

\bibitem{VlasovPetrishchevTalanovRdiofiz1971}
S.~N. Vlasov, V.~A. Petrishchev, and V.~I. Talanov,
\enquote{Averaged
  description of wave beams in linear and nonlinear media,} Izv. Vys. Uchebn.
  Zaved. Radiofizika \textbf{14}, 1353 (1971).

\bibitem{ZakharovJETP1972}
V.~E. Zakharov, \enquote{Collapse of langmuir waves,} Sov. Phys.
JETP
  \textbf{35}, 908 (1972).

\bibitem{SulemSulem1999}
C.~Sulem and P.~L. Sulem, \emph{Nonlinear Schr\"odinger Equations:
  Self-Focusing and Wave Collapse} (World Scientific, New York, 1999).

\bibitem{LushnikovDyachenkoVladimirovaNLSloglogPRA2013}
P.~M. Lushnikov, S.~A. Dyachenko, and N.~Vladimirova,
\enquote{{Beyond
  leading-order logarithmic scaling in the catastrophic self-focusing of a
  laser beam in Kerr media},} Phys. Rev. A \textbf{88}, 013845 (2013).

\bibitem{LushnikovRosePRL2004}
P.~M. Lushnikov and H.~A. Rose, \enquote{Instability versus
equilibrium
  propagation of laser beam in plasma,} Phys. Rev. Lett. \textbf{92}, 255003
  (2004).

\bibitem{LushnikovRosePlasmPysContrFusion2006}
P.~M. Lushnikov and H.~A. Rose, \enquote{How much laser power can
propagate
  through fusion plasma?} Plasma Physics and Controlled Fusion \textbf{48},
  1501--1513 (2006).

\bibitem{LushnikovVladimirovaOptLett2010}
P.~M. Lushnikov and N.~Vladimirova, \enquote{Non-gaussian statistics
of
  multiple filamentation,} Opt. Lett. \textbf{35}, 1965--1967 (2010).

\bibitem{ChungLushnikovPRE2011}
Y.~Chung and P.~M. Lushnikov, \enquote{Strong collapse turbulence in
quintic
  nonlinear schr\"odinger equation.} Physical Review E \textbf{84}, 036602
  (2011).

\bibitem{RossBookLaserBeamQualityMetrics2013}
T.~S. Ross, \emph{Laser Beam Quality Metrics} (SPIE Press, 2013).

\bibitem{SmithDoApplOpt2008}
A.~V. Smith and B.~T. Do, \enquote{Bulk and surface laser damage of
silica by
  picosecond and nanosecond pulses at 1064 nm,} Applied Optics \textbf{47},
  4812--4832 (2008).

\bibitem{DawsonALLOptExpr2008}
J.~W. Dawson, M.~J. Messerly, R.~J. Beach, Miroslav, Y.~Shverdin,
E.~A.
  Stappaerts, A.~K. Sridharan, P.~H. Pax, J.~E. Heebner, C.~W. Siders, and
  C.~Barty, \enquote{Analysis of the scalability of diffraction-limited fiber
  lasers and amplifiers to high average power,} Optics Communications
  \textbf{16}, 13240--13266 (2008).

\end{thebibliography}
\end{document}